\documentclass[11pt]{article}

\usepackage[final]{acl}

\usepackage[T1]{fontenc}

\usepackage[utf8]{inputenc}
\usepackage{microtype}

\usepackage{times}
\usepackage{latexsym}
\usepackage{amsmath}
\usepackage{amssymb}
\usepackage{bbm}
\usepackage{makecell}
\usepackage{multirow}
\usepackage{kotex}
\usepackage{array}
\usepackage{booktabs}
\usepackage{graphicx}
\usepackage{tcolorbox}
\usepackage{xcolor}
\usepackage{colortbl}
\usepackage{tabularx}

\usepackage{algorithm}
\usepackage{algpseudocode}

\title{MIMO: Multilingual Information Retrieval via Monolingual Objectives}

\author{
    Youngjoon Jang, Seongtae Hong, Heuiseok Lim\thanks{Corresponding author} \\
    Department of Computer Science and Engineering, Korea University \\
    \texttt{\{dew1701, ghdchlwls123, limhseok\}@korea.ac.kr}
}

\begin{document}
\maketitle
\begin{abstract}

Multilingual Information Retrieval (MLIR) reflects real-world search environments in which queries and relevant documents may appear in different languages within a mixed-language corpus. However, existing embedding models are primarily optimized for Multi-Monolingual retrieval and their performance often degrades in MLIR settings. Moreover, directly applying conventional contrastive learning to MLIR can exacerbate language clustering and expose a trade-off between cross-lingual alignment and embedding uniformity. To address these limitations, we propose \textbf{MIMO}: \textbf{M}ultilingual \textbf{I}nformation Retrieval via \textbf{M}onolingual \textbf{O}bjectives, 
a two-stage framework that uses a stable English semantic space from a high-performing teacher model as an anchor. MIMO first initializes the student model's cross-lingual alignment through knowledge distillation, and then jointly optimizes distillation and cross-lingual contrastive learning to improve retrieval discrimination while preserving alignment. Extensive experiments show that MIMO consistently outperforms existing cross-lingual training baselines across various MLIR and Multi-Monolingual benchmarks. MIMO also remains competitive with off-the-shelf models of similar or larger parameter scales. Furthermore, our cross-lingual Alignment-Uniformity analysis clarifies the distinct roles of the two loss components and shows that their combination yields a favorable trade-off between alignment and uniformity. The code for \textbf{MIMO} is available at \url{https://github.com/yjoonjang/MIMO}.
\end{abstract}

\section{Introduction}

As Information Retrieval (IR) technologies become globally ubiquitous, real-world user search environments are inevitably becoming multilingual~\citep{hong1996multilingual, peters2012multilingual, sujatha2011review, rahimi2015multilingual}. For instance, even when an English speaker issues a query in English, the most relevant and high-quality documents may be written in other languages, such as Korean, Chinese, or Spanish. This open-ended environment, where user queries and target documents can be in any arbitrary, distinct languages, is defined as Multilingual Information Retrieval (MLIR)~\citep{peters2002importance}. To maximize user accessibility to information in the global web ecosystem, search systems capable of effectively handling these MLIR scenarios are essential~\citep{nie2000multilingual, si2017empirical}.

\begin{figure}[t!]
\centering
\includegraphics[width=1.0\linewidth]{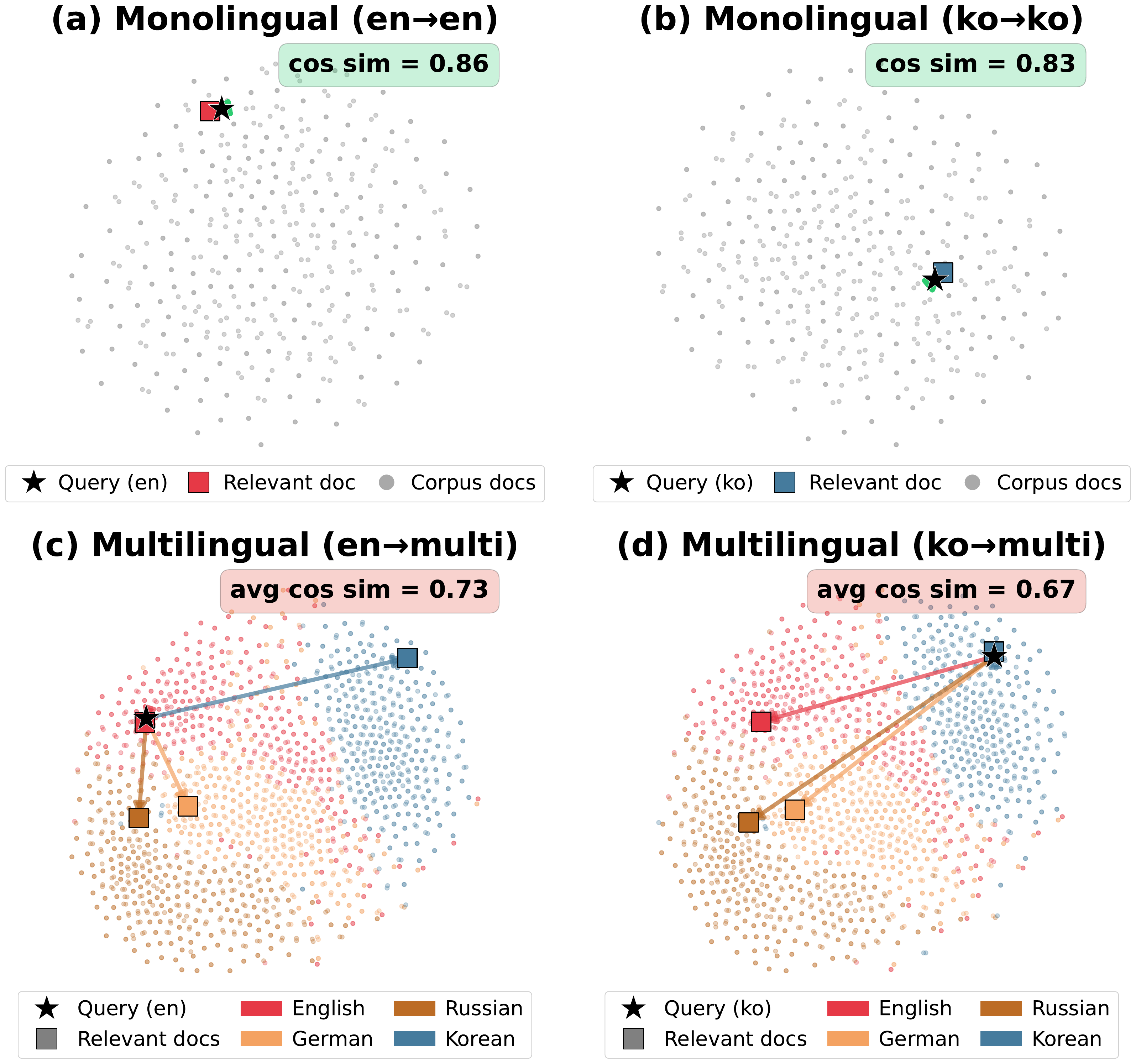}
\caption{An illustration of the embedding space using xlm-roberta-large model trained with standard InfoNCE loss on monolingual pairs on the Belebele parallel dataset. (a) and (b) show monolingual retrieval scenarios where queries and corpus are in the same language. (c) and (d) demonstrate multilingual scenarios where corpus consist of mixed languages.}
\label{fig:problem_statement}
\end{figure}

Recently, advancements in dense retrieval based on multilingual Pre-trained Language Models (mPLMs) have significantly improved the performance of models handling multiple languages~\citep{zhang2024mgte, chen2024bge, akram2026jina, zhang2025qwen3, eslami2026diffusion}. However, existing multilingual retrieval models, along with their training strategies and evaluation datasets, are predominantly optimized for Multi-Monolingual IR scenarios, where the query and document languages remain identical. While these models exhibit outstanding matching performance within individual languages, they reveal severe limitations in MLIR scenarios that require identifying relevant answers across language boundaries within a mixed-language corpus~\citep{park2025investigating, hong2026improving}. 

The issues faced by conventional training methodologies in MLIR environments are clearly observed in Figure~\ref{fig:problem_statement}. This figure visualizes queries and documents from the Belebele parallel dataset~\citep{bandarkar2024belebele} in the embedding space, using an xlm-roberta-large encoder fine-tuned exclusively with the InfoNCE objective on monolingual query-positive pairs. As shown in (a) and (b), in Multi-Monolingual scenarios where the query and document languages are identical, the model successfully places the query and its positive document close together, yielding high cosine similarities of 0.86 and 0.83. 

Conversely, (c) and (d) illustrate MLIR scenarios where a single query is evaluated against a mixed-language corpus. Specifically, the model is tasked with retrieving from a mixed corpus where four semantically identical documents are translated into four different languages. In these cases, the data points are severely clustered by language rather than semantics. Even though the four documents possess identical meanings, their superficial linguistic differences cause the average cosine similarity to drop significantly to 0.73 and 0.67. This indicates that InfoNCE, a widely-used training method, fails to achieve proper cross-lingual alignment, leading to the formation of language-specific islands and ultimately causing a degradation in MLIR performance.

To resolve these issues, we propose MIMO (Multilingual Information Retrieval via Monolingual Objectives). MIMO utilizes a robust English embedding space from a high-performing teacher model as an absolute anchor. The student model initializes its cross-lingual alignment space through a Stage 1 knowledge distillation warmup, learning to map multilingual inputs into the teacher's English semantic space. In the subsequent Stage 2, it jointly optimizes knowledge distillation and cross-lingual contrastive learning. Notably, while it was known from \citet{wang2020understanding} that contrastive learning improves embedding uniformity, our study indicates that knowledge distillation drives the alignment of the embedding space in cross-lingual environments. Building upon this discovery, MIMO effectively separates the roles of each loss function, providing an effective operating point to balance the alignment-uniformity trade-off.
In summary, the main contributions of this paper are as follows:
\begin{itemize}
    \item We propose MIMO, a two-stage training pipeline for MLIR that uses a stable English semantic anchor provided by a teacher model to combine knowledge distillation and cross-lingual contrastive learning for multilingual representation learning.
    \item We extend the Alignment-Uniformity analysis to cross-lingual settings, clarifying the distinct roles of the distillation and contrastive objectives and showing that their combination provides a favorable trade-off between alignment and uniformity.
    \item MIMO consistently outperforms existing cross-lingual training baselines on MLIR and multi-monolingual benchmarks. It also achieves highly competitive performance against off-the-shelf models of similar or larger parameter scales.
\end{itemize}

\section{Related Works}
\label{sec:relwork}
\subsection{Multilingual Retrieval Settings}
Multilingual retrieval environments can be defined in various ways depending on the language composition of the query and the corpus. Building upon the definitions discussed in the early Cross-Language Evaluation Forum (CLEF) and NeuCLIRBench, we categorize the environments of modern retrieval benchmarks as shown in Table \ref{tab:ir_settings}~\citep{braschler2002clef, lawrie2025neuclirbench}.

\begin{table}[h]
\centering
\resizebox{\columnwidth}{!}{%
\begin{tabular}{lccc}
\hline
\textbf{Setting} & \textbf{Query Language} & \textbf{Corpus Language} \\
\hline
Monolingual IR & $L_1$ (fixed) & $L_1$ (fixed) \\
Cross-Lingual IR (CLIR) & $L_1$ (fixed) & $L_2$ (fixed) \\
Multi-Monolingual IR & $L$ & $L$ (same, but $L$ varies) \\
\textbf{Multilingual IR (MLIR)} & $L$ (any) & $L_{mixed}$  \\
\hline
\end{tabular}%
}
\caption{Comparison of multilingual retrieval settings. $L$, $L_1$, and $L_2$ denote specific languages.}
\label{tab:ir_settings}
\end{table}

Despite MLIR being the most realistic setting for global web search, recent benchmarks predominantly focus on Multi-Monolingual IR, where query and document languages are paired identically~\citep{zhang-etal-2021-mr, chen2024bge, akram2026jina, eslami2026diffusion}. Even advanced mixed-corpus benchmarks like NeuCLIRBench restrict queries strictly to English~\citep{lawrie2025neuclirbench}, leaving a critical gap in evaluating any-to-any language retrieval.

\subsection{Training Methods}

\paragraph{Contrastive Learning}
The InfoNCE loss is the standard dense retrieval objective, optimizing models by pulling a query closer to a positive document while pushing away in-batch negatives~\citep{karpukhin2020dense}. However, existing multilingual retrieval models predominantly train on monolingual query-document pairs~\citep{babakhin2025llamaembednemotron8buniversaltextembedding}. Therefore, models trained solely with this formulation possess a fundamental limitation in effectively bridging the semantic gap between texts written in different languages.

\paragraph{Cross-Lingual Alignment}
To enhance cross-lingual alignment, \citet{chi2021infoxlm} proposed a Cross-lingual Contrastive (XLCO) Objective that directly utilizes query-document pairs in different languages. In a similar vein, \citet{yang-etal-2024-language-bias} proposed the LaKDA loss, which minimizes the Symmetric KL Divergence between the similarity probability distributions of semantically parallel query pairs over a batch of documents. However, these methodologies share a common limitation: they remain confined to relative alignment, attempting to match multilingual representations internally within the student model. Due to the absence of a stable absolute anchor, there is a distinct limit to constructing the unified high-dimensional semantic space required for high absolute retrieval performance in an MLIR environment.

\paragraph{Knowledge Distillation}
To introduce an absolute anchor, Knowledge Distillation has been actively explored. Notably, \citet{reimers2020making} constructed multilingual sentence embeddings by encouraging a student model to mimic the embedding space of a proven English monolingual model using Mean Squared Error. While highly effective in cross-lingual transfer, its evaluation is largely confined to Multi-Monolingual scenarios and Semantic Textual Similarity (STS) tasks. 

Building upon this, our MIMO framework utilizes a robust English embedding space as an absolute anchor. We advance this approach through a novel two-stage training strategy and extend the Alignment-Uniformity analysis~\citep{wang2020understanding} to cross-lingual environments. We empirically demonstrate that while pure distillation is highly effective at forming cross-lingual alignment, it can compromise embedding uniformity, the core component of Information Retrieval required to discriminate subtle semantic differences among vast amounts of documents. MIMO effectively mitigates this trade-off via joint optimization, achieving an optimal balance between robust alignment and spatial uniformity.

\section{MIMO}

We propose MIMO, a novel two-stage training framework for MLIR environments. MIMO adopts a Knowledge Distillation architecture that designates a high-performance embedding model as the teacher and a relatively smaller multilingual language model as the student. Specifically, the teacher model observes only inputs in English throughout the training process, providing a constant and stable semantic anchor. The student model receives inputs of the same meaning in various languages and learns to project them into the English embedding space of the teacher model. Through this, it effectively aligns multilingual representations into a unified semantic space.

\subsection{Stage 1: Cross-Lingual Distillation}
The objective of the first stage of MIMO is to initialize the student model's multilingual embedding space to mimic the English embedding space of the teacher model. This encourages the student model to acquire foundational cross-lingual understanding capabilities, mapping texts from various languages into a shared semantic space prior to undertaking full-scale retrieval-specific training.

\paragraph{Dataset}
For Stage 1 training, we utilize English-Multilingual text translation pairs ($t_{en}, t_{XX}$) from the OPUS parallel corpus~\citep{tiedemann-2012-parallel, zhang-etal-2020-improving}. Specifically, we select the same 14 target languages used in the mMARCO dataset~\citep{bonifacio2021mmarco} for Stage 2 training, sampling up to 500,000 sentence pairs per language to ensure training balance. Details are provided in Appendix~\ref{app:stage1_data}.

\paragraph{Training}
The teacher model $E_t$ encodes the corresponding English text $t_{en}$, while the student model $E_s$ encodes the text $t_{XX}$ composed in an arbitrary language. To match the embedding dimensions between the two models, we apply a linear projection layer $\psi: \mathbb{R}^n \rightarrow \mathbb{R}^m$ to the output of the student model, following \citet{kim2023embeddistill}. Here, $n$ and $m$ are the hidden dimension sizes of the student and teacher models respectively, and $\psi(z)$ is defined as $\psi(z) = Wz + b$. 
Finally, training proceeds to minimize the cosine distance between the two representations~\citep{wang2018cosface}:
\begin{equation*}
\mathcal{L}_{Stage1} = 1 - \cos(\psi(E_s(t_{XX})), E_t(t_{en}))
\end{equation*}

\subsection{Stage 2: Joint Optimization for MLIR}
Knowledge distillation alone is limited in addressing the Information Retrieval task, which requires identifying the correct answer among vast candidate documents. Therefore, Stage 2 employs a training approach that combines Distillation and Contrastive Learning. 

\paragraph{Dataset}
We utilize the mMARCO parallel dataset comprising 14 languages to jointly optimize cross-lingual contrastive learning, which acquires retrieval capabilities, and knowledge distillation, which retains the teacher model's knowledge. Each training data instance consists of an English query $q_{en}$ and a positive passage $p_{en}$, along with their corresponding translated query $q_{XX}$ and positive passage $p_{YY}$ in arbitrary languages: $\{q_{en}, p_{en}, q_{XX}, p_{YY}\}$. The details are provided in Appendix~\ref{app:stage2_data}.

\paragraph{Embedding Distillation Loss}
The distillation loss is calculated by reusing the same linear projection layer $\psi$ from Stage 1. The teacher model encodes the English query $q_{en}$ and document $p_{en}$ to provide a stable target space that the student model must reach:
\begin{equation*}
\begin{split}
\mathcal{L}_{Distill} &= [1 - \cos(\psi(E_s(q_{XX})), E_t(q_{en}))] \\
&\quad + [1 - \cos(\psi(E_s(p_{YY})), E_t(p_{en}))]
\end{split}
\end{equation*}
This loss prevents the embedding space from fragmenting by language and serves as an anchor to ensure that the student model retains the rich semantic properties transferred from the teacher model. 

\paragraph{Cross-lingual Contrastive Loss}
To secure the student model's retrieval capabilities, we also include a cross-lingual contrastive learning loss (XLCO). The student model encodes pairs of query $q_{XX}$ and document $p_{YY}$ composed in arbitrary, distinct languages, and the contrastive loss is calculated using a temperature parameter $\tau$ and in-batch negatives $p_j$ within a batch size $B$ as follows:
\begin{equation*}
\mathcal{L}_{XLCO} = - \log \frac{\exp(s(q_{XX}, p_{YY}) / \tau)}{\sum_{j=1}^{B} \exp(s(q_{XX}, p_j) / \tau)}
\end{equation*}
This loss function distances the positive document from numerous negative documents, training the model to acquire the discrimination ability essential for retrieval tasks.

\paragraph{Joint Optimization Objective}
Finally, the Stage 2 training loss of MIMO is formulated as the weighted sum of these two components:
\begin{equation*}
\mathcal{L}_{Stage2} = \lambda \cdot \mathcal{L}_{XLCO} + (1 - \lambda) \cdot \mathcal{L}_{Distill}
\end{equation*}
The weight parameter $\lambda$ plays a key role in balancing the acquisition of retrieval discrimination ($\mathcal{L}_{XLCO}$) and the retention of cross-lingual knowledge ($\mathcal{L}_{Distill}$).

\section{Experimental Setup}
\subsection{Training Configuration}
We use xlm-roberta-large~\citep{conneau2020unsupervisedcrosslingualrepresentationlearning} and mmBERT-base~\citep{marone2025mmbertmodernmultilingualencoder} with mean pooling as our student backbones, and Qwen3-Embedding-8B~\citep{zhang2025qwen3} as the teacher model. To enable the large-scale in-batch negatives essential for contrastive learning, we employ gradient caching~\citep{gao2021scalingdeepcontrastivelearning} to achieve a global batch size of 2048. We set the Stage 2 loss weight $\lambda$ to 0.2. Comprehensive training details are provided in Appendix~\ref{app:training_data} and Appendix~\ref{app:training_details}.

\subsection{Evaluation Benchmarks}
To evaluate MIMO in MLIR environments, we utilize six distinct evaluation benchmarks.
\begin{table*}[t!]
\centering
\resizebox{1.0\textwidth}{!}{%
\begin{tabular}{cl|cccccc||c}
\toprule
\textbf{Model} & \multicolumn{1}{c|}{\textbf{Method}} & \textbf{Belebele (14)} & \textbf{MLQA (7)} & \textbf{XQuAD (8)} & \textbf{MultiEuP (6)} & \textbf{NeuCLIR22 (2)} & \textbf{NeuCLIR23 (2)} & \textbf{AVG} \\ \midrule

\multirow{4}{*}{xlm-roberta-large}
 & InfoNCE & 35.86 & 24.86 & 42.47 & 17.23 & 36.83 & 31.20 & 31.41 \\ 
 & XLCO & 74.02 & 41.95 & 77.50 & 41.02 & 46.99 & 37.62 & 53.18 \\
 & LaKDA & 73.90 & 41.99 & 77.54 & 41.09 & 46.20 & 36.96 & 52.95 \\
 & \textbf{MIMO (Ours)} & \textbf{84.18} & \textbf{49.83} & \textbf{86.05} & \textbf{47.06} & \textbf{55.40} & \textbf{41.62} & \textbf{60.69} \\ \midrule

\multirow{4}{*}{mmBERT-base}
 & InfoNCE & 31.45 & 25.86 & 39.92 & 15.61 & 40.19 & 32.12 & 30.86 \\
 & XLCO & 70.70 & 40.21 & 76.00 & 35.97 & 45.28 & 39.34 & 51.25 \\
 & LaKDA & 70.79 & 40.27 & 75.94 & 36.08 & 44.84 & 39.93 & 51.31 \\
 & \textbf{MIMO (Ours)} & \textbf{80.27} & \textbf{46.06} & \textbf{83.18} & \textbf{45.11} & \textbf{50.51} & \textbf{43.37} & \textbf{58.08} \\ \bottomrule
\end{tabular}%
}
\caption{MLIR performance comparison of the proposed MIMO method against baselines. The numbers in parentheses indicate the number of languages evaluated for each dataset. Best results are highlighted in bold.}
\label{tab:main}
\end{table*}

\paragraph{MLIR}
We first construct an MLIR benchmark by combining high-quality parallel datasets: Belebele~\citep{bandarkar2024belebele}, MLQA~\citep{lewis2020mlqa}, MultiEuP-v2~\cite{yang-etal-2024-language-bias}, and XQuAD~\citep{Artetxe:etal:2019}. 
While these datasets inherently contain queries and semantically linked documents across multiple languages, they are conventionally evaluated in individual language pairs. To adapt these datasets for MLIR, we construct a unified multilingual retrieval corpus by mixing all language versions of the context passages, inspired by LAReQA~\citep{roy2020lareqa}. For each query, we treat the source-language gold context and all target-language versions of that context as positive documents, as they share completely identical semantics but differ only in language. In this setup, the model must successfully retrieve all relevant multilingual positive documents from the mixed corpus, regardless of language. In our experiments, we utilize the subset of 14 languages that overlap with mMARCO. 

Furthermore, we additionally utilize the NeuCLIR 2022~\citep{lawrie2023overview} and NeuCLIR 2023~\citep{lawrie2024overview} benchmarks, which are widely used for evaluating existing MLIR. These benchmarks require retrieving answers from a mixed corpus of Chinese (zh), Russian (ru), and Persian (fa) given an English (en) query. For fair evaluation, we filter the corpus to include only Chinese and Russian documents, as these languages overlap with our mMARCO training data. Following the convention of prior studies, we use nDCG@20 as the evaluation metric for all MLIR settings~\citep{lawrie2025neuclirbench}. 

\paragraph{Multi-Monolingual IR} 
Additionally, to evaluate Multi-Monolingual IR performance, we conduct evaluations on 10 languages (overlapping with mMARCO) using the MIRACL benchmark, a widely used standard for evaluating multilingual retrieval models. Unlike MLIR, this setting represents a retrieval environment where the languages of the query and the search corpus are fixed to be identical. According to the official specifications, nDCG@10 is used as the evaluation metric.

All evaluations are conducted using MTEB\footnote{\url{https://github.com/embeddings-benchmark/mteb}}~\citep{muennighoff2022mteb, enevoldsen2025mmtebmassivemultilingualtext}. Detailed descriptions and statistics of each dataset used for evaluation can be found in Appendix~\ref{app:eval_data}.

\subsection{Baselines}
To evaluate the effectiveness of the MIMO methodology, we select representative loss functions widely used in training retrievers (InfoNCE, XLCO, LaKDA) as our baselines, as discussed in Section~\ref{sec:relwork}. To ensure experimental rigor, all baseline models are trained under exactly the same conditions as MIMO (identical backbone models, datasets, hyperparameters, and computing resources), differing only in the formulation of the loss function. Detailed implementation settings for the baselines are provided in Appendix~\ref{app:baselines}.

Additionally, to provide a broader context against established production models, we evaluate several strong off-the-shelf multilingual retrievers in a zero-shot setting. A comprehensive comparison and analysis of these models are detailed in Appendix~\ref{app:off_the_shelf}.

\section{Results \& Analysis}

\subsection{Main Results}
\label{sec:main}

Table \ref{tab:main} presents the MLIR performance (nDCG@20) evaluated on six multilingual retrieval benchmarks based on two student models. 

Experimental results show that the proposed MIMO framework consistently and substantially outperforms all existing training baselines across both models. Notably, InfoNCE, which utilizes only monolingual query-document pairs, exhibited the lowest performance. This indicates that contrastive learning alone tends to cluster embeddings by language, leading to failures in a mixed-language corpus. While XLCO and LaKDA, which utilize cross-lingual pairs, improved performance over InfoNCE, they still fall short of MIMO. This suggests that combining knowledge distillation using the teacher model's English embedding space with contrastive learning is a robust and generalized training methodology that endows multilingual information retrieval capabilities, regardless of parameter scale.

\subsection{Impact of Training Stages}
\begin{table}[ht!]
\centering
\resizebox{\columnwidth}{!}{%
\begin{tabular}{l|ccc}
\toprule
\textbf{Dataset} & \textbf{Stage1 Only} & \textbf{Stage2 Only} & \textbf{Stage1 + Stage2} \\ \midrule
Belebele  & 83.22 & 73.29 & \textbf{84.39} \\
MLQA  & 45.26 & 41.10 & \textbf{49.86} \\
XQuAD  & \textbf{87.02} & 76.90 & 86.41 \\
MultiEuP  & 40.70 & 40.24 & \textbf{46.99} \\
NeuCLIR22  & 48.50 & 46.83 & \textbf{55.40} \\
NeuCLIR23 & 39.88 & 38.29 & \textbf{41.62} \\ \midrule
\textbf{AVG}  & 57.43 & 52.77 & \textbf{60.78} \\ \bottomrule
\end{tabular}%
}
\caption{MLIR performance (nDCG@20) comparison across different training stages.}
\label{tab:stage_impact}
\end{table}

MIMO consists of a two-stage training pipeline: Knowledge Distillation Warmup (Stage 1) and Joint Optimization (Stage 2). Table \ref{tab:stage_impact} presents the ablation study results to verify the effectiveness of this two-stage approach. All experiments are conducted using the xlm-roberta-large backbone model.

When only the Stage 1 warmup is performed (Stage 1 Only), the student model merely learns to mimic the English embeddings from the teacher rather than the retrieval task itself, and therefore underperforms the full pipeline. Nevertheless, it even surpasses XLCO and LaKDA in Table~\ref{tab:main}. This suggests that Teacher Distillation alone already provides strong cross-lingual alignment.

Furthermore, when the Stage 2 joint optimization loss is applied directly to the backbone without warmup (Stage 2 Only), the average retrieval performance is in fact the lowest among the three settings, falling below even the distillation-only model. This demonstrates that, prior to cross-lingual contrastive fine-tuning, it is highly beneficial to first initialize the student model's multilingual representations into the teacher model's English embedding space to form a foundational alignment; without this warmup, contrastive learning struggles to organize the multilingual representation space.

\subsection{Optimal Balance of Loss Components}

\begin{figure}[h!]
\centering
\includegraphics[width=1.0\linewidth]{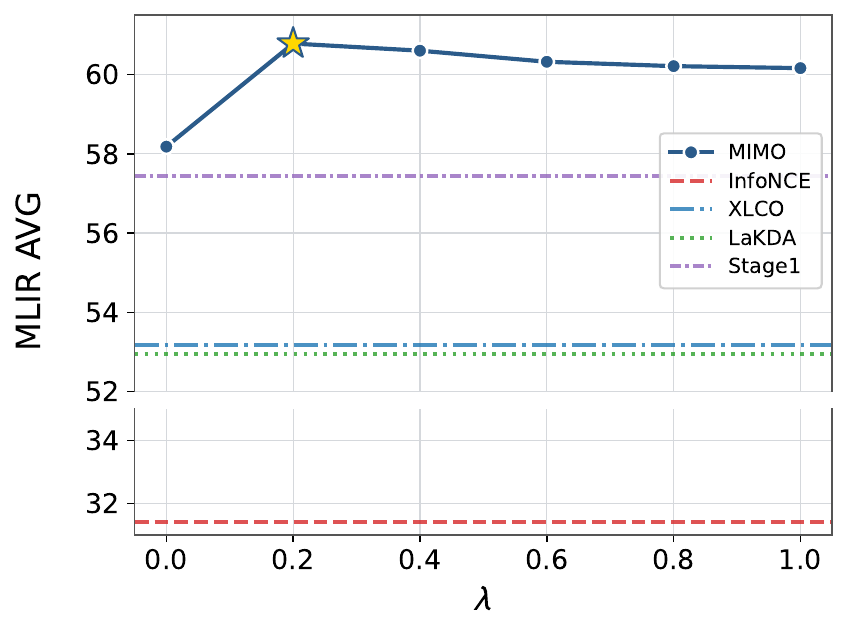}
\caption{Impact of the weight parameter $\lambda$ on MLIR average performance (nDCG@20). Horizontal lines denote the performance of the baselines and the Stage 1 warmup model. $\lambda=0.0$ represents pure distillation ($\mathcal{L}_{Distill}$), while $\lambda=1.0$ represents pure cross-lingual contrastive learning ($\mathcal{L}_{XLCO}$).}
\label{fig:lambda_ablation}
\end{figure}

In MIMO's Stage 2 joint loss function, the weight parameter $\lambda$ determines the balance between contrastive learning ($\mathcal{L}_{XLCO}$) and knowledge distillation ($\mathcal{L}_{Distill}$). Figure \ref{fig:lambda_ablation} shows the average MLIR performance according to varying values of $\lambda$, alongside the baseline performances for context.
Experimental results show that pure Knowledge Distillation ($\lambda=0.0$) records the lowest score among the variants, and pure XLCO ($\lambda=1.0$) shows relatively high performance. This implies that learning discrimination ability through cross-lingual contrastive learning is necessary for the retrieval task.

Crucially, MIMO consistently outperforms all existing baselines across the entire range of $\lambda$. The highest performance is achieved at $\lambda=0.2$, establishing a substantial gap over the strongest baselines (XLCO and LaKDA). Performance steadily decreases as $\lambda$ increases beyond this point, indicating that the optimal operating point is achieved when blending the two losses with a higher proportion of Distillation. This suggests that while the discrimination power provided by contrastive learning is essential for retrieval tasks, an excessive proportion can compromise the rich knowledge distilled from the teacher model.

\subsection{Alignment-Uniformity Analysis}
\label{sec:A_U}
To diagnose the underlying mechanics across varying $\lambda$, we extend the Alignment-Uniformity framework \citep{wang2020understanding} to the MLIR domain. We evaluate this using the fully parallel Belebele dataset across 14 languages, redefining positive pairs as all 14 semantically identical documents in different languages. 

\begin{figure}[h!]
\centering
\includegraphics[width=1.0\linewidth]{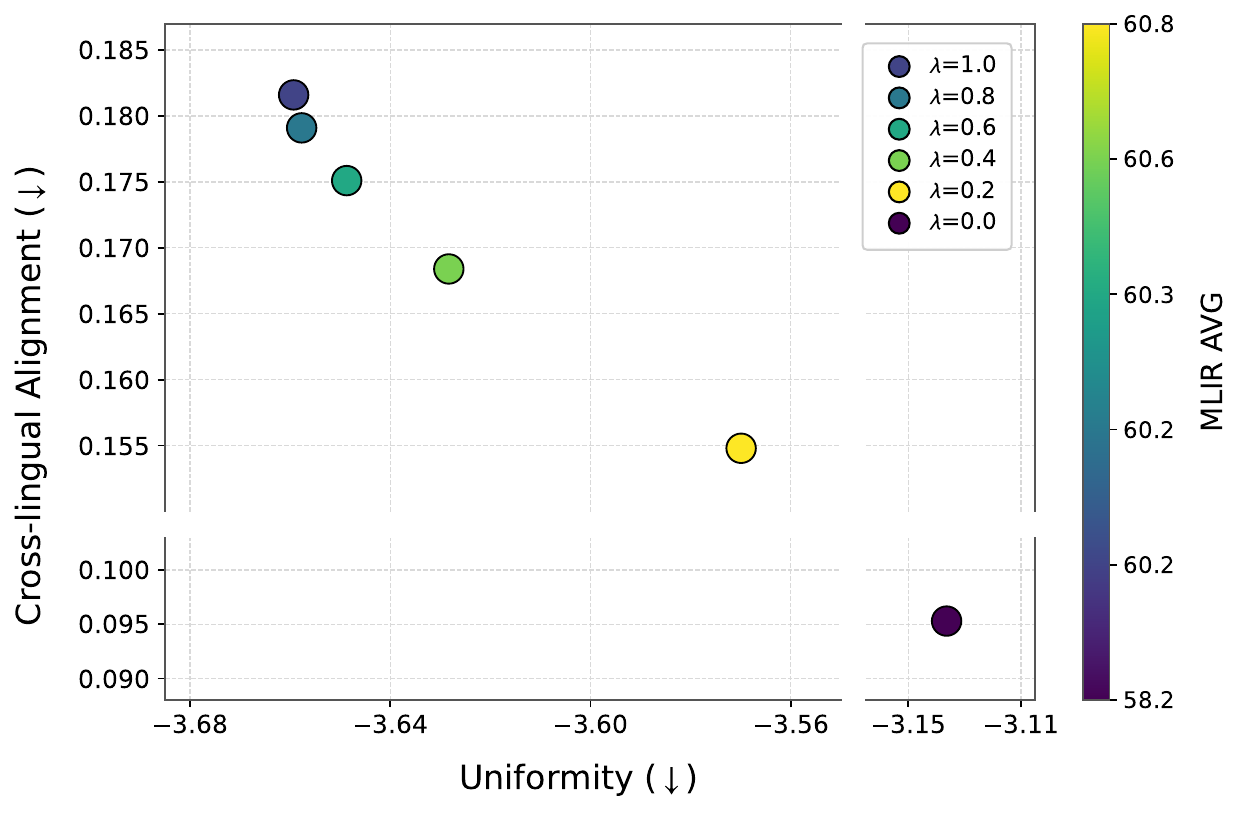}
\caption{Alignment and Uniformity analysis of MIMO variants with different $\lambda$ values. Lower values indicate better cross-lingual alignment and better uniformity.}
\label{fig:alignment_uniformity}
\end{figure}

\paragraph{Metrics} 
Cross-lingual Alignment measures the mean squared distance between strictly parallel cross-lingual documents:
\begin{equation*}
\ell_{\text{align}} = \mathbb{E}_{(x_{L_i}, x_{L_j})} \| f(x_{L_i}) - f(x_{L_j}) \|^2
\end{equation*}
Here, $f(\cdot)$ is the encoder, and $(x_{L_i}, x_{L_j})$ denotes a pair of semantically identical documents translated into distinct languages $L_i$ and $L_j$. Crucially, this expectation is computed across all possible language pairs among the 14 Belebele languages, ensuring an evaluation of the universal semantic space. Lower $\ell_{\text{align}}$ values indicate that identical semantics are closely matched regardless of language. 
Conversely, Uniformity calculates the pairwise Gaussian potential:
\begin{equation*}
\ell_{\text{uniform}} = \log \mathbb{E}_{(x, y)} \, e^{-t\|f(x)-f(y)\|^2}
\end{equation*}
Here, $(x, y)$ represents any two documents independently sampled from the entire mixed 14-language corpus, regardless of their language or meaning, and $t$ is a hyperparameter (set to $t=2$ in our experiment). Lower $\ell_{\text{uniform}}$ values signify evenly distributed embeddings, securing the discriminative power essential for retrieval.

\paragraph{Analysis of the Trade-off}
Figure \ref{fig:alignment_uniformity} confirms that the two loss components perform distinct roles, forming a clear trade-off. 

Pure knowledge distillation ($\lambda=0.0$) yields the best (lowest) alignment by strongly binding multilingual representations to the teacher's anchor. However, it severely compromises spatial uniformity, losing the discrimination power needed to distinguish among numerous negative documents. Conversely, pure contrastive learning ($\lambda=1.0$) maximizes (scores lowest) uniformity but noticeably degrades cross-lingual alignment. 

Ultimately, MIMO's joint optimization at $\lambda=0.2$ finds an optimal balance in this trade-off. It secures sufficient spatial uniformity for robust retrieval while preserving much of the deep cross-lingual alignment transferred from the teacher. For a broader Alignment-Uniformity comparison against existing training baselines, we provide additional analysis in Appendix~\ref{app:additional_alignment_uniformity}.

\subsection{Language Fairness and Consistency}
\begin{figure}[h!]
\centering
\includegraphics[width=1.0\linewidth]{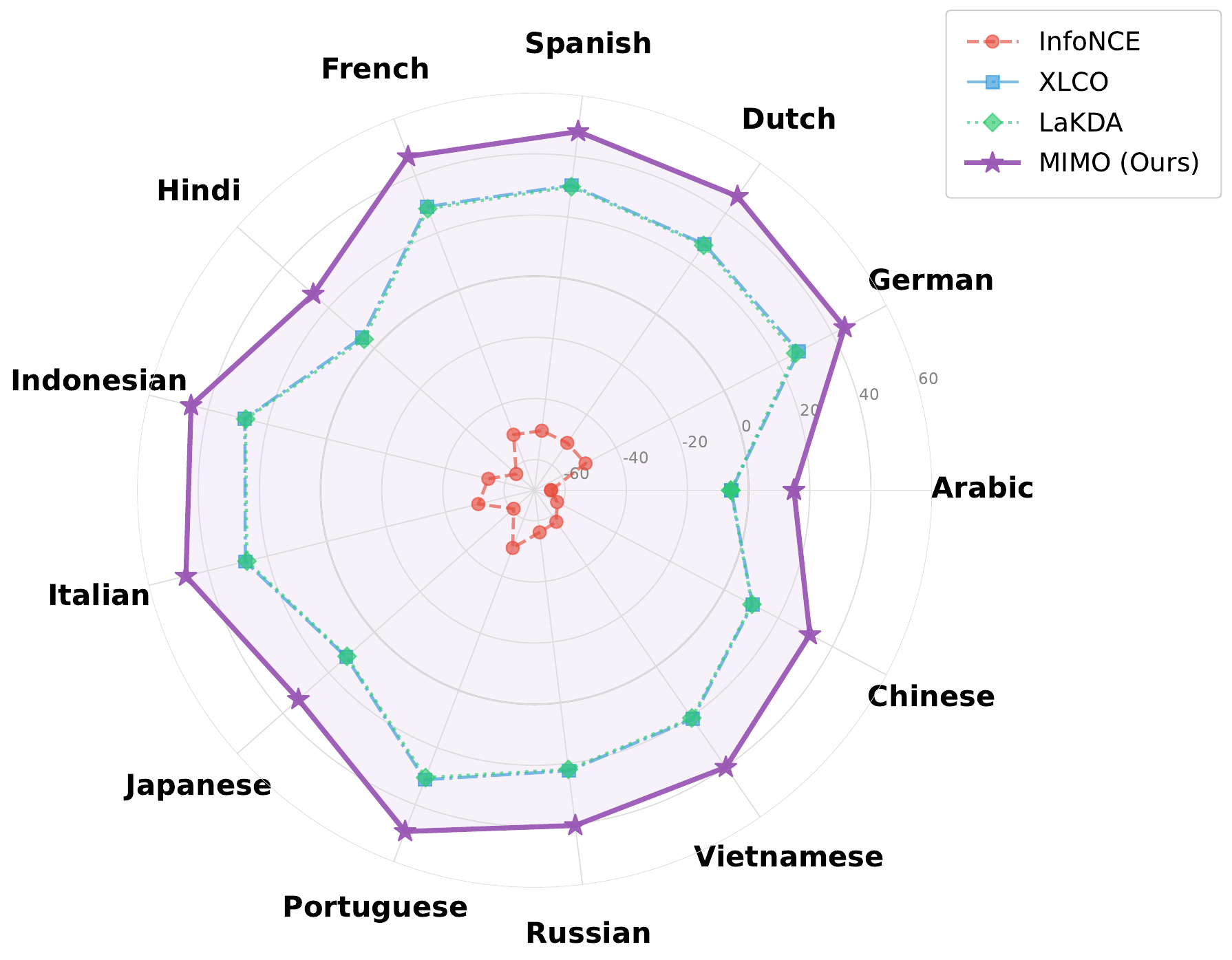}
\caption{Radar chart of Multilingual Rank-based Consistency (MRC@10) across different query languages. The scores are scaled by a factor of 100.}
\label{fig:mrc_radar}
\end{figure}

For a MLIR system to be trusted in a real-world environment, it must provide consistent retrieval results without language bias, regardless of the language in which the user inputs the query. To evaluate the linguistic fairness of the models, we adopt the Multilingual Rank-based Consistency (MRC) metric recently proposed by \citet{yang-etal-2024-language-bias}. MRC evaluates whether the document rankings returned by the retrieval model remain consistent when semantically identical queries are input in different languages.

\begin{table*}[t!]
\centering
\resizebox{\textwidth}{!}{
\begin{tabular}{ll|cccccccccc||c}
\toprule
\textbf{Model} & \textbf{Method} & \textbf{ar} & \textbf{de} & \textbf{en} & \textbf{es} & \textbf{fr} & \textbf{hi} & \textbf{id} & \textbf{ja} & \textbf{ru} & \textbf{zh} & \textbf{AVG} \\
\midrule
\multirow{4}{*}{xlm-roberta-large} 
& InfoNCE & 51.88 & 33.75 & 35.74 & 39.20 & 33.61 & 36.86 & 37.88 & 39.01 & 38.81 & 36.44 & 38.32 \\
& XLCO    & 51.34 & 37.95 & 38.04 & 39.46 & 35.13 & 37.49 & 38.15 & 40.41 & 40.65 & 41.28 & 39.99 \\
& LaKDA   & 52.40 & 38.56 & 38.41 & 39.72 & 36.96 & 37.98 & 38.41 & 41.10 & 41.70 & 41.09 & 40.63 \\
& \textbf{MIMO}    & \textbf{53.40} & \textbf{43.81} & \textbf{42.72} & \textbf{43.10} & \textbf{43.46} & \textbf{41.19} & \textbf{39.51} & \textbf{44.11} & \textbf{47.54} & \textbf{42.78} & \textbf{44.16} \\
\midrule
\multirow{4}{*}{mmBERT-base} 
& InfoNCE & \textbf{50.26} & 35.16 & 35.73 & 39.73 & 33.75 & \textbf{36.15} & 36.40 & \textbf{40.58} & 36.65 & 34.29 & 37.87 \\
& XLCO    & 44.23 & 37.54 & 37.42 & 38.78 & 35.91 & 29.94 & 36.66 & 39.71 & 36.17 & 38.26 & 37.46 \\
& LaKDA   & 44.87 & 37.59 & 37.57 & 39.26 & 36.21 & 30.76 & \textbf{37.37} & 39.95 & 36.48 & 38.40 & 37.85 \\
& \textbf{MIMO}    & 43.65 & \textbf{40.28} & \textbf{40.79} & \textbf{41.38} & \textbf{39.09} & 31.78 & 36.56 & 40.26 & \textbf{38.88} & \textbf{42.46} & \textbf{39.51} \\
\bottomrule
\end{tabular}
}
\caption{Multi-Monolingual performance comparison across different models and methods.}
\label{tab:tb_miracl}
\end{table*}

\paragraph{Measurement of MRC}
Specifically, the MRC@$k$ metric is calculated using a parallel query set composed of $L=14$ languages from the Belebele dataset. For a specific query index $i$, let $R(q_{(i,a)})$ denote the top-$k$ document ranking returned for the query in target language $a$. The pairwise consistency between language $a$ and another language $b$ is computed using Spearman's rank correlation, $\text{corr}(R(q_{(i,a)}), R(q_{(i,b)}))$, over the union of their retrieved documents. The consistency score for query $i$ in language $a$ is obtained by averaging these correlation coefficients across all other $L-1$ languages:
\begin{equation}
RC_{(a)}^i = \frac{1}{L-1} \sum_{b \neq a} \text{corr}(R(q_{(i,a)}), R(q_{(i,b)}))
\end{equation}
The final MRC@$k$ score for a particular language $a$ is then calculated by averaging these consistency scores across all $N$ test queries:
\begin{equation}
\text{MRC}@k_{(a)} = \frac{1}{N} \sum_{i=1}^N RC_{(a)}^i
\end{equation}
In short, a higher MRC score and a radar chart closer to a regular polygon suggest that retrieval performance is more stable and consistent across query languages, indicating lower sensitivity to the superficial language used for the query.

\paragraph{Language Bias in Existing Methods}
Figure \ref{fig:mrc_radar} visualizes the MRC@10 scores per query language for the four training methodologies. The InfoNCE model, which relies solely on monolingual query-document pairs, exhibits severe language bias where ranking consistency is compromised depending on the query language. XLCO, which introduces cross-lingual contrastive learning, and LaKDA, which combines a distribution alignment loss specifically designed to ensure linguistic fairness, improved overall consistency compared to InfoNCE. However, they still exhibit irregular shapes with consistency scores fluctuating depending on the query language.

\paragraph{Superior Consistency of MIMO}
In contrast, the proposed MIMO framework consistently outperforms the existing baseline models across all query languages, forming a broader and more uniform trajectory. Notably, MIMO's consistency, which utilizes the teacher's target space, is more stable than the LaKDA method, which explicitly sets cross-lingual alignment and language bias mitigation as its direct training objectives. This shows that MIMO's two-stage pipeline, combining knowledge distillation and contrastive learning, strongly binds multilingual representations to the teacher's semantic space, thereby constructing a more robust retrieval system unswayed by the superficial linguistic form of the query.

\subsection{Multi-Monolingual Retrieval Performance}

While the primary objective of this study is to improve the MLIR environment, it is also crucial to verify whether the proposed model can maintain and enhance robust performance in traditional retrieval settings. To this end, we conduct additional evaluations using MIRACL, a Multi-Monolingual retrieval benchmark where the languages of the query and the search corpus remain identical.

Table~\ref{tab:tb_miracl} presents the evaluation results on the MIRACL benchmark across 10 languages. Experimental results show that MIMO achieves the highest average performance, substantially outperforming existing baselines on average for both backbone models. InfoNCE, based on monolingual contrastive learning, recorded the lowest average score, followed by XLCO and LaKDA. MIMO established a substantial average performance gap over LaKDA, the highest-performing baseline.

These results suggest that MIMO's two-stage joint optimization framework is not merely a localized methodology addressing only the Cross-lingual Alignment problem. It indicates that the high-quality semantic knowledge transferred from the English embedding space of the teacher model enriches the overall representational power of the student model and improves the uniformity of the embedding space, thereby advancing query-document matching capabilities even within the same language.

\section{Conclusion}
In this paper, we propose MIMO, a two-stage training framework for Multilingual Information Retrieval (MLIR). By anchoring to a high-performing teacher's English semantic space, MIMO combines knowledge distillation and cross-lingual contrastive learning to achieve an optimal balance between cross-lingual alignment and embedding uniformity. MIMO consistently outperforms existing training baselines across both MLIR and Multi-Monolingual benchmarks. Furthermore, its strong MRC results demonstrate that MIMO delivers stable, unbiased retrieval performance regardless of the query language.

\section*{Limitations}
While the proposed MIMO framework demonstrates substantial improvements in multilingual retrieval and language fairness, this study has a few limitations that point to directions for future work. 

First, our training and evaluation are currently confined to 14 languages that overlap with the mMARCO dataset. Although these languages cover a diverse range of language families, further research is required to scale the framework to a much broader set of languages (e.g., 100+ languages). Expanding the evaluation to include extremely low-resource languages is necessary to fully verify the universal applicability of the proposed methodology.

Second, the training pipeline of MIMO fundamentally relies on parallel corpora. Specifically, parallel sentence pairs for the Stage 1 warmup and translated query-document pairs for the Stage2 joint optimization. While such parallel data is becoming increasingly accessible, the strict requirement for translated pairs can act as a bottleneck when extending the model to underrepresented languages lacking high-quality translation resources. Exploring advanced methods to reduce this dependency, such as leveraging unsupervised or weakly supervised cross-lingual alignment techniques, remains an important objective for future research.

Finally, we note that our constructed MLIR benchmarks are based on parallel corpora, which inherently rewards exact translation-level alignment. Real-world multilingual web retrieval may involve less strict parallelism and more topical diversity.




\section*{Ethics Statement}
This study utilizes publicly available datasets and open-source libraries for all experiments. We adhered to the licenses and terms of use associated with these resources. Regarding the use of AI tools, we utilized AI assistants (ChatGPT, Gemini) exclusively for grammatical error correction and polishing the text to enhance readability. The scientific ideas and contributions presented in this paper are entirely our own.

\section*{Acknowledgments}
This research was supported by Basic Science Research Program through the National Research Foundation of Korea(NRF) funded by the Ministry of Education(NRF-2021R1A6A1A03045425).
This work was supported by Institute for Information \& communications Technology Promotion(IITP) grant funded by the Korea government(MSIT) 
(RS-2024-00398115, Research on the reliability and coherence of outcomes produced by Generative AI).
This work was supported by the Commercialization Promotion Agency for R\&D Outcomes(COMPA) grant funded by the Korea government(Ministry of Science and ICT)(2710096072).
This work was supported by the Institute of Information \& Communications Technology Planning \& Evaluation(IITP)-ICT Creative Consilience Program grant funded by the Korea government(MSIT)(IITP-2026-RS-2020-II201819).

\bibliography{custom}

\clearpage

\appendix

\section{Baselines}
\label{app:baselines}

We compare MIMO against three representative training strategies for multilingual retrieval.

\paragraph{InfoNCE}
The standard contrastive learning objective using monolingual query-positive pairs $(q_L, p_L)$:
\begin{equation*}
\mathcal{L}_{\text{InfoNCE}} = -\log \frac{\exp(s(q_L, p_L)/\tau)}
{\sum_{j=1}^{B} \exp(s(q_L, p_j)/\tau)}
\end{equation*}
where $s(\cdot,\cdot)$ denotes cosine similarity and $\tau = 1/20$
is the temperature in our experiment. Since query and document share the same language,
this baseline lacks any explicit cross-lingual learning signal.

\paragraph{XLCO}
Cross-lingual Contrastive (XLCO) Objective uses cross-lingual pairs $(q_{XX}, p_{YY})$ where $XX \neq YY$, applying the same InfoNCE formulation but with the query and document in different languages. This forces the model to align representations across language boundaries through contrastive learning alone.
\begin{equation*}
\mathcal{L}_{XLCO} = -\log \frac{\exp(s(q_{XX}, p_{YY}) / \tau)}{\sum_{p_j \in B} \exp(s(q_{XX}, p_j) / \tau)}
\end{equation*}

\paragraph{LaKDA}
Language-aware Knowledge Distillation and Alignment (LaKDA) combines contrastive learning with cross-lingual distribution alignment. Given a query $q$, the probability distribution $P(p_j|q)$ over documents $p_j$ within batch $B$ is defined as:
\begin{equation*}
P(p_j|q) = \frac{\exp(s(q, p_j)/\tau)}{\sum_{p_k \in B} \exp(s(q, p_k)/\tau)}
\end{equation*}
Letting $\mathbf{P}_1$ and $\mathbf{P}_2$ be the distributions derived from two semantically parallel queries $q_{L_1}$ and $q_{L_2}$ (the same query expressed in different languages), LaKDA minimizes a weighted combination of contrastive loss and Symmetric KL Divergence:
\begin{equation*}
\begin{split}    
\mathcal{L}_{\text{LaKDA}} = & \; (1-\alpha) \cdot \mathcal{L}_{\text{InfoNCE}} \\
& + \alpha \cdot \frac{1}{2} \Big[ D_{\text{KL}}(\mathbf{P}_1 \| \mathbf{P}_2)
+ D_{\text{KL}}(\mathbf{P}_2 \| \mathbf{P}_1) \Big]
\end{split}
\end{equation*}
Following the original implementation, we set $\alpha = 0.5$ in our experiment. 

\section{Training Dataset Details}
\label{app:training_data}

\subsection{Stage 1: Cross-lingual Distillation}
\label{app:stage1_data}

For the Stage~1 knowledge distillation warmup, we construct a parallel sentence dataset from OPUS~\footnote{\url{https://opus.nlpl.eu/}}, which aggregates eight multilingual parallel corpora: Europarl, GlobalVoices, JW300, News-Commentary, OpenSubtitles, TED Talks, Tatoeba, and WikiMatrix. Each training instance is an English sentence paired with its translation, forming the pair $\{t_\text{en}, t_\text{XX}\}$. We apply quality filters to remove noisy pairs: minimum length of 10 characters, maximum length of 512 characters, and a length ratio (longer/shorter) below 3:1. English-side deduplication is performed to prevent duplicate anchors across languages. Each language is capped at 500,000 pairs to limit over-representation of high-resource languages, resulting in a total of 5,647,936 training pairs across 14 languages. Table~\ref{tab:stage1_data} summarizes the per-language statistics.

\begin{table}[t]
\centering
\resizebox{\columnwidth}{!}{%
\begin{tabular}{lrlr}
\toprule
\textbf{Language} & \textbf{\# of Pairs} & \textbf{Language} & \textbf{\# of Pairs} \\
\midrule
Arabic (ar)     & 500,000 & Indonesian (id) & 500,000 \\
German (de)     & 500,000 & Italian (it)    & 500,000 \\
Dutch (nl)      & 500,000 & Japanese (ja)   & 192,283 \\
English (en)    & 500,000 & Portuguese (pt) & 500,000 \\
Spanish (es)    & 500,000 & Russian (ru)    & 500,000 \\
French (fr)     & 500,000 & Vietnamese (vi) & 278,507 \\
Hindi (hi)      & 141,739 & Chinese (zh)    &  35,407 \\
\midrule
\multicolumn{2}{l}{\textbf{Total}} & \multicolumn{2}{r}{\textbf{5,647,936}} \\
\bottomrule
\end{tabular}
}
\caption{Stage~1 parallel sentence data statistics per language.}
\label{tab:stage1_data}
\end{table}

\subsection{Stage 2: Joint Optimization for MLIR}
\label{app:stage2_data}

For Stage~2 joint optimization, we utilize the mMARCO parallel dataset~\citep{bonifacio2021mmarco}, a multilingual translation of the MSMARCO passage retrieval corpus covering 14 languages. The resulting dataset contains 415,938 \{query, positive\} pairs across 14 languages: Arabic (ar), German (de), Dutch (nl), English (en), Spanish (es), French (fr), Hindi (hi), Indonesian (id), Italian (it), Japanese (ja), Portuguese (pt), Russian (ru), Vietnamese (vi), and Chinese (zh).
To ensure fair language representation, we employ a uniform round-robin sampling strategy with per-cycle shuffling. 

\paragraph{Data Composition for MIMO \& XLCO}
For MIMO and XLCO, which use cross-lingual pairs $(q_{XX}, p_{YY})$ where $XX \neq YY$, all 182 ($14 \times 13 = 182$) ordered language pairs appear with equal frequency (2,286 times each, max-min difference of 1), resulting in equal 415,938 pairs. 

\paragraph{Data Composition for InfoNCE}
For InfoNCE, which uses monolingual pairs, each of the 14 language pairs appear 29,710 times. This uniform distribution prevents any language pair from dominating the training signal. 

\paragraph{Data Composition for LaKDA}
For LaKDA, which requires two query languages $(q_{XX}, q_{YY})$ with $XX \neq YY$ and an independent positive passage $p$,
the query language pairs $(XX, YY)$ follow the same uniform distribution as MIMO (182 pairs, 2,286 times each),
while the positive passage language is sampled independently with uniform distribution across 14 languages (29,710 times each).


\section{Training Details}
\label{app:training_details}

\paragraph{Teacher Model}
We use Qwen3-Embedding-8B~\cite{zhang2025qwen3} as the teacher model, which achieves top-ranking performance on MTEB~\cite{muennighoff2022mteb} at the time of our experiments. The teacher model has approximately 8 billion parameters and produces 3,584 dimensional embeddings. It is loaded in bfloat16 precision with Flash Attention~2 and remains completely frozen (no gradient computation) throughout training.

\paragraph{Teacher Prompt}
Following the Qwen3-Embedding specification, the teacher model uses the following query prompt during Stage~2 encoding:
\begin{quote}
\small
\texttt{Instruct: Given a web search query, retrieve relevant passages that answer the query\textbackslash nQuery: }
\end{quote}
No prompt is applied to document inputs. The student model does not use any prompt during training or inference. Also, we note that we do not use any prompt in Stage~1 training.

\paragraph{Projection Layer}
Since the student and teacher models have different embedding dimensions (1,024 for xlm-roberta-large, 768 for mmBERT-base, and 3,584 for the teacher), we introduce a learnable linear projection $\psi: \mathbb{R}^{d_s} \rightarrow \mathbb{R}^{d_t}$ to map student embeddings into the teacher's space. This projection is initialized and trained during Stage~1, then used again in Stage~2. Importantly, the projection is discarded at inference time.

\paragraph{Gradient Caching}
To achieve the large batch size of 2,048 required for effective contrastive learning, we employ the gradient cache technique~\cite{gao2021scalingdeepcontrastivelearning}. Forward passes are computed in mini-batches of 32 samples, and gradients are accumulated and propagated after all mini-batches are processed. This enables the use of batch sizes that would otherwise exceed GPU memory.

\paragraph{Hyperparameters}
Table~\ref{tab:hyperparams} summarizes the hyperparameters for each training stage.

\begin{table}[t]
\centering
\small
\begin{tabular}{lcc}
\toprule
\textbf{Hyperparameter} & \textbf{Stage 1} & \textbf{Stage 2} \\
\midrule
Optimizer           & AdamW     & AdamW \\
Learning rate       & $1 \times 10^{-4}$ & $2 \times 10^{-5}$ \\
Warmup ratio        & 0.1       & 0.1 \\
Batch size (per GPU)& 128       & 2,048 \\
Max seq length (student)  & 256 & 512 \\
Max seq length (teacher)  & 256 & 512 \\
Epochs              & 1         & 1 \\
Mixed Precision           & bfloat16  & bfloat16 \\
GPUs                & 2$\times$ H100 & 2$\times$ H100 \\
$\lambda$ (contrastive weight) & --- & 0.2 \\
InfoNCE scale ($1/\tau$) & --- & 20.0 \\
\bottomrule
\end{tabular}
\caption{Hyperparameters for Stage~1 and Stage~2 training.}
\label{tab:hyperparams}
\end{table}

\section{Evaluation Benchmarks}
\label{app:eval_data}

All evaluations are conducted using MTEB~\cite{muennighoff2022mteb}. Table~\ref{tab:eval_stats} provides a comprehensive summary of all evaluation datasets.

\subsection{MLIR Benchmarks}
\label{app:mmlir}

We evaluate MLIR performance on four benchmarks where queries in any language must retrieve relevant documents from a mixed multilingual corpus. For all MLIR benchmarks, we use nDCG@20 as the evaluation metric following the convention of prior studies~\cite{lawrie2025neuclirbench}.

\paragraph{Belebele} 
Belebele is a multilingual machine reading comprehension dataset spanning 122 language variants. Based on the FLORES-200 benchmark, each instance consists of a short passage and a corresponding multiple-choice question. For the retrieval task, we utilize the question stem as the query and the passage as the positive document. A defining feature of Belebele is its full parallelism: the entire set of questions and passages represents strictly parallel content across all 122 languages.
For our evaluation, we select the 14 languages that overlap with the mMARCO dataset: Arabic (ar), German (de), Dutch (nl), English (en), Spanish (es), French (fr), Hindi (hi), Indonesian (id), Italian (it), Japanese (ja), Portuguese (pt), Russian (ru), Vietnamese (vi), and Chinese (zh).

\paragraph{XQuAD} 
XQuAD is a cross-lingual QA benchmark consisting of a subset of the SQuAD v1.1 development set. The dataset was constructed by translating both the questions and the paragraphs from English into 10 target languages. This rigorous human translation process guarantees that XQuAD is fully parallel; every question and paragraph has an exact, semantically aligned counterpart in all other languages. 
We utilize the 8 languages intersecting with mMARCO: Arabic (ar), German (de), English (en), Spanish (es), Hindi (hi), Russian (ru), Vietnamese (vi), and Chinese (zh).

\paragraph{MLQA} 
MLQA is a multi-way aligned extractive QA benchmark covering seven languages. Unlike Belebele or XQuAD, MLQA is designed to be multi-way parallel; each QA instance is aligned across a subset of languages (typically four) rather than the entire set. This structure results from mining parallel sentences where overlaps exist across varying language combinations to maximize linguistic diversity. Consequently, queries are also not fully parallel across all seven languages for every instance. Instead, for a given document, queries exist only in the specific subset of languages aligned with that document. Despite this partial parallelism, the dataset ensures that within each 4-way subset, the questions and contexts are semantically equivalent.
Since all 7 languages overlap with mMARCO, we utilize the language set for evaluation: Arabic (ar), German (de), English (en), Spanish (es), Hindi (hi), Vietnamese (vi), and Simplified Chinese (zh).

\paragraph{MultiEup-v2} 
Following \citet{yang-etal-2024-language-bias}, we utilize MultiEup-v2 as an MLIR benchmark. The task is defined as retrieving the relevant parliamentary speech segment (document) given a subject descriptor (query). The dataset is constructed from European Parliament proceedings, where professional archivists manually assign official subject descriptors (e.g., `International Human Rights') to specific speech segments. Since the descriptors are professionally translated into 24 languages, the same information can be expressed as multilingual queries. In this framework, the ground truth is the specific speech text that was originally tagged with the descriptor by human experts.
We use the 6 languages that overlap with mMARCO: German (de), English (en), Spanish (es), French (fr), Italian (it), and Portuguese (pt).

\paragraph{NeuCLIR}
We evaluate on NeuCLIR 2022 and NeuCLIR 2023 under a true MLIR setting, in which a single English query must retrieve relevant documents from a corpus that mixes multiple languages, following the multilingual retrieval definition of NeuCLIRBench~\citep{lawrie2025neuclirbench}. Constructing this setting requires care, because the MTEB NeuCLIR hard-negatives subsets~\footnote{\url{https://huggingface.co/datasets/mteb/NeuCLIR2022RetrievalHardNegatives}}\footnote{\url{https://huggingface.co/datasets/mteb/NeuCLIR2023RetrievalHardNegatives}} are organized per language. 
To make this to a MLIR setting, we (i) take the original English (en) queries of NeuCLIRBench~\footnote{\url{https://huggingface.co/datasets/neuclir/bench}}, (ii) pool the Chinese (zh) and Russian (ru) hard-negatives corpora into a single mixed collection, excluding Persian (fas) as it lies outside our 14 training languages, and (iii) judge relevance using the union of the per-language cross-lingual relevance judgments (\texttt{qrels.rus} $\cup$ \texttt{qrels.zho}) provided by NeuCLIRBench, matched to each English query by its identifier. We keep NeuCLIR 2022 and 2023 as separate tracks, since their query sets are disjoint, and report nDCG@20 consistent with the other MLIR benchmarks.

We note that Belebele, XQuAD, MLQA, NeuCLIR (and its hard negatives mined versions) are all utilized in MMTEB~\cite{enevoldsen2025mmtebmassivemultilingualtext}, while MultiEup-v2 is adopted to evaluate MLIR tasks following~\citet{yang-etal-2024-language-bias}. 

\begin{table}[t]
\centering
\resizebox{0.8\columnwidth}{!}{%
\begin{tabular}{lrrc}
\toprule
\textbf{Dataset} & \textbf{\# Queries} & \textbf{\# Corpus} & \textbf{\# Langs} \\
\midrule
Belebele   & 12,600 &  6,832 & 14  \\
MLQA       & 42,245 & 36,799 &  7  \\
XQuAD      &  9,520 &  1,920 &  8  \\
MultiEuP   & 10,686 & 35,444 &  6  \\
NeuCLIR'22 &     49 & 19,049 & 2  \\
NeuCLIR'23 &     63 & 33,512 & 2 \\
\bottomrule
\end{tabular}
}
\caption{Evaluation benchmark statistics.}
\label{tab:eval_stats}
\end{table}

\subsection{Multi-Monolingual Benchmark}
\label{app:miracl}

For Multi-Monolingual IR evaluation, we use the MIRACL benchmark~\citep{zhang-etal-2023-miracl} with hard negatives version~\footnote{\url{https://huggingface.co/datasets/mteb/MIRACLRetrievalHardNegatives}} across 10 languages: Arabic (ar), Chinese (zh), English (en), French (fr), German (de), Hindi (hi), Indonesian (id), Japanese (ja), Russian (ru), and Spanish (es). Unlike MLIR, this setting evaluates retrieval where the query and corpus share the same language. We use nDCG@10 as the evaluation metric following the official benchmark specifications.

\begin{table*}[t!]
\centering
\resizebox{\textwidth}{!}{%
\begin{tabular}{l c| cccccc|| c}
\toprule
\textbf{Model} & \textbf{Size} & \textbf{Belebele} & \textbf{MLQA} & \textbf{XQuAD} & \textbf{MultiEuP} & \textbf{NeuCLIR22} & \textbf{NeuCLIR23} & \textbf{AVG} \\
\midrule
\multicolumn{9}{l}{\textit{Large Models ($\ge$ 1B)}} \\
\midrule
Qwen3-Embedding-8B (Teacher) & 8B & 96.29 & 69.28 & 95.02 & 50.88 & 68.87 & 55.35 & 72.62 \\
gte-Qwen2-1.5B-Instruct & 1.5B & 56.71 & 30.39 & 56.45 & 8.70 & 33.42 & 19.48 & 34.19 \\
\midrule
\multicolumn{9}{l}{\textit{Small Models ($<$ 1B)}} \\
\midrule
Qwen3-Embedding-0.6B & 0.6B & 81.16 & 55.88 & 86.52 & 43.10 & 59.35 & 48.19 & 62.37 \\
bge-m3 & 0.6B & 84.02 & 57.33 & 89.89 & 43.77 & 43.19 & 40.03 & 59.71 \\
nomic-embed-text-v2-moe & 0.5B & 78.52 & 53.82 & 84.66 & 43.90 & 47.67 & 46.22 & 59.13 \\
multilingual-e5-large-instruct & 0.6B & 40.76 & 35.81 & 48.60 & 24.27 & 55.52 & 49.28 & 42.37 \\
multilingual-e5-large & 0.6B & 61.51 & 43.01 & 69.98 & 30.87 & 9.69 & 18.55 & 38.93 \\
\midrule
\textbf{MIMO (Ours)} & \textbf{0.6B} & \textbf{84.18} & \textbf{49.83} & \textbf{86.05} & \textbf{47.06} & \textbf{55.40} & \textbf{41.62} & \textbf{60.69} \\
\bottomrule
\end{tabular}%
}
\caption{MLIR performance (nDCG@20) comparison against strong off-the-shelf multilingual embedding models. Models are grouped by their parameter scale. We highlight the xlm-roberta-large model trained with MIMO.}
\label{tab:off_the_shelf}
\end{table*}

\section{Comparison with Off-the-Shelf Models}
\label{app:off_the_shelf}

To provide a broader context for the performance of MIMO, we evaluate several off-the-shelf multilingual embedding models in a zero-shot setting on our MLIR benchmarks. 

As shown in Table~\ref{tab:off_the_shelf}, MIMO operates within the small models category (0.6B scale). Despite its relatively small size, MIMO achieves a highly competitive average score of 60.69. Notably, it substantially outperforms both multilingual-e5-large and multilingual-e5-large-instruct, which share the exact same backbone architecture~\citep{wang2024multilingual}. Furthermore, it surpasses both the nomic-embed-text-v2-moe~\citep{nussbaum2025trainingsparsemixtureexperts}, bge-m3~\citep{chen2024bge} and the much larger 1.5B gte-Qwen2-1.5B-Instruct~\citep{li2023towards}. It is particularly noteworthy that off-the-shelf models like the nomic-embed-text-v2-moe and bge-m3 rely heavily on extensive hard negative mining pipelines to achieve their performance. In stark contrast, MIMO achieves this highly competitive performance utilizing only simple in-batch negatives on standard parallel corpora, without requiring any complex hard negative mining. This highlights the inherent strength and alignment capability of the proposed training objective.

Predictably, MIMO trails slightly behind heavily engineered production models like Qwen3-Embedding-0.6B~\citep{zhang2025qwen3} within the same size category. We emphasize that this remaining performance gap stems from vast differences in engineering scale rather than the fundamental training objective. Specifically, Qwen3-Embedding-0.6B is trained on massive, proprietary datasets, incorporating billions of carefully mined hard-negative pairs and complex multi-stage pipelines. In contrast, MIMO was trained on a highly restricted set of datasets. 

Therefore, while the Qwen3-Embedding-0.6B model represent the upper bound of current industry capabilities, MIMO's performance demonstrates the orthogonal value of our proposed loss formulation. Integrating the MIMO objective into such large-scale production pipelines could yield even stronger cross-lingual alignment in future research.

\section{Additional Alignment-Uniformity Analysis}
\label{app:additional_alignment_uniformity}

\begin{figure}[h!]
\centering
\includegraphics[width=1.0\linewidth]{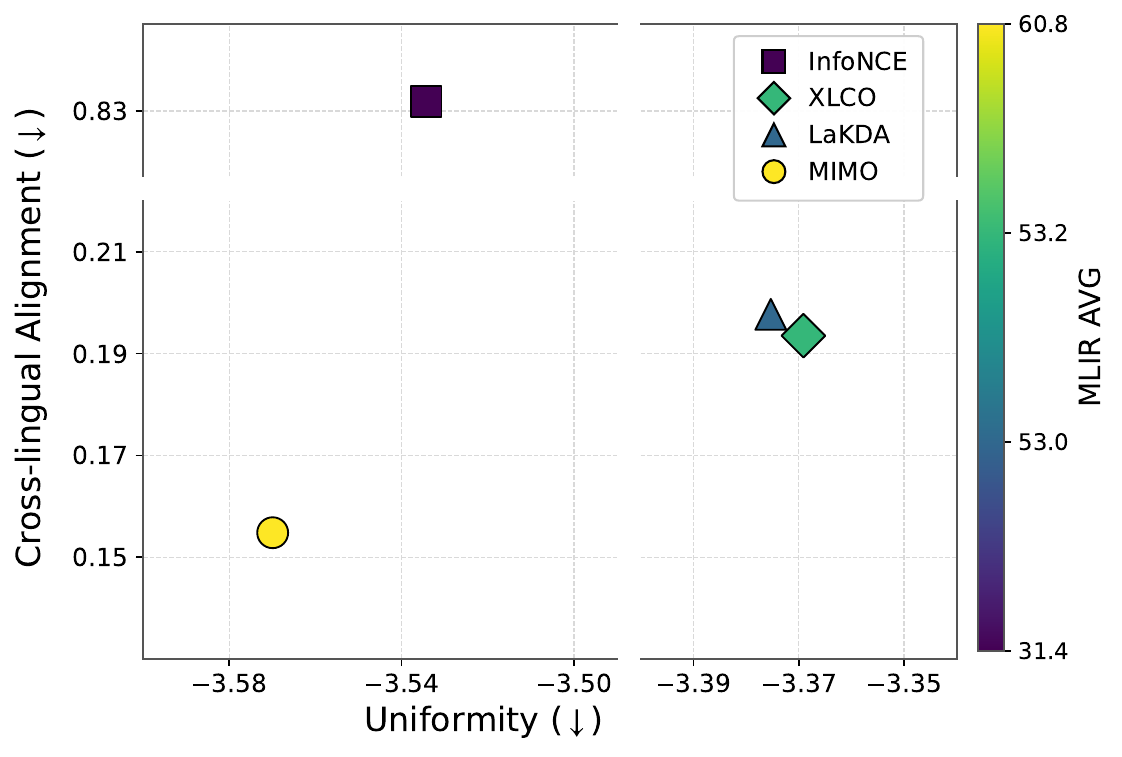}
\caption{Alignment and Uniformity analysis of baselines and MIMO. Lower values indicate better cross-lingual alignment and better uniformity.}
\label{fig:alignment_uniformity_appen}
\end{figure}

\begin{table*}[t!]
\centering
\resizebox{1.0\textwidth}{!}{%
\begin{tabular}{cl|cccccc||c}
\toprule
\textbf{Model} & \multicolumn{1}{c|}{\textbf{Method}} & \textbf{Belebele (14)} & \textbf{MLQA (7)} & \textbf{XQuAD (8)} & \textbf{MultiEuP (6)} & \textbf{NeuCLIR22 (2)} & \textbf{NeuCLIR23 (2)} & \textbf{AVG} \\ \midrule

\multirow{4}{*}{\begin{tabular}[c]{@{}c@{}}mmBERT-base\\(cls)\end{tabular}}
 & InfoNCE & 29.85 & 24.92 & 38.03 & 15.04 & 39.21 & 33.87 & 30.15 \\
 & XLCO & 67.89 & 38.51 & 71.94 & 34.63 & 40.13 & 34.41 & 47.92 \\
 & LaKDA & 68.13 & 38.67 & 72.21 & 34.87 & 40.72 & 34.87 & 48.24 \\
 & \textbf{MIMO (Ours)} & \textbf{78.48} & \textbf{44.61} & \textbf{81.24} & \textbf{44.72} & \textbf{51.19} & \textbf{41.78} & \textbf{57.01} \\ \bottomrule
\end{tabular}%
}
\caption{MLIR performance comparison using the mmBERT-base model with [CLS] pooling.}
\label{tab:mmbert_cls}
\end{table*}

In Section~\ref{sec:A_U}, we analyzed the Alignment-Uniformity trade-off strictly within the variations of MIMO's joint optimization weight $\lambda$. To provide a broader context, Figure~\ref{fig:alignment_uniformity_appen} extends this analysis to compare MIMO against the baseline training methodologies: InfoNCE, XLCO, and LaKDA. The results vividly demonstrate the fundamental limitations of existing baselines and the superiority of the MIMO framework.

\paragraph{InfoNCE fails in cross-lingual alignment}
The InfoNCE model, trained solely on monolingual pairs, exhibits an excessively poor cross-lingual alignment loss (approximately $0.83$). This visualizes the phenomenon discussed in Figure~\ref{tab:ir_settings}, where the model completely fails to bridge the semantic gap across different languages, clustering solely by superficial linguistic forms. Consequently, it records the lowest MLIR average performance.

\paragraph{Trade-off in XLCO and LaKDA}
Methodologies explicitly utilizing cross-lingual pairs, such as XLCO and LaKDA, successfully reduce the alignment loss to around 0.19. However, this comes at a severe cost to uniformity, which degrades substantially (shifting rightwards to approximately -3.37). This indicates that while relative alignment techniques pull parallel documents together, their embedding spaces severely collapse. They lose the spatial uniformity strictly necessary to discriminate among a vast pool of negative documents, trapping them in a sub-optimal trade-off.

\paragraph{MIMO balances alignment and uniformity}
In stark contrast, MIMO successfully breaks this trade-off barrier. By anchoring to the teacher's stable English semantic space and jointly optimizing the objectives, MIMO secures the most optimal position in the bottom-left quadrant. It achieves the best (lowest) cross-lingual alignment ($\sim$0.15) while maintaining excellent uniformity ($\sim$-3.57). This ideal geometric structure of the embedding space translates directly to the highest MLIR performance (60.78), establishing a massive gap over all existing baselines.

\section{Variant of Pooling Strategy}
\label{sec:appendix_pooling}

In Section~\ref{sec:main}, we employed mean pooling to derive sequence-level embeddings from the student backbones. To verify that the effectiveness of the proposed MIMO framework is robust and not strictly dependent on a specific pooling method, we conduct an additional evaluation using the [CLS] token representation.

Table \ref{tab:mmbert_cls} presents the MLIR performance of the mmBERT-base model trained and evaluated using [CLS] pooling. The experimental results demonstrate that MIMO consistently maintains its superiority over all baseline methodologies. Specifically, MIMO achieves an average nDCG@20 of 57.01, establishing a substantial gap of 8.77\%p over LaKDA, the strongest baseline in this setting.

While the absolute average performance of [CLS] pooling (57.01) is slightly lower than that of mean pooling (58.09), the relative performance trends across the training methods remain identical. This confirms that MIMO's joint optimization strategy fundamentally improves MLIR performance regardless of the underlying pooling mechanism used.

\end{document}